\RequirePackage[running]{lineno}

\documentclass[aps,prd,onecolumn,nofootinbib,superscriptaddress,letterpaper,amsmath,amssymb]{revtex4-2}
\pdfoutput=1

\usepackage{ulem}
\usepackage{graphicx}   % needed for figures
\usepackage{dcolumn}    % needed for some tables
\usepackage{bbm}        % for math
\usepackage{amsmath}
\usepackage{amsfonts}
\usepackage{amssymb}   % for math
\usepackage{bm}
\usepackage{epstopdf}
\usepackage{slashed}
\usepackage{hyperref}
\usepackage{multirow}
\usepackage{color}
\usepackage{xcolor} 
\usepackage{float}
\usepackage{subcaption}

\hypersetup{hidelinks}

\newcommand{\xspec}{{\sc xspec }}

\newcommand{\surrEOS}{ML-Likelihood$_{\text{EOS}}$}

\definecolor{highlight}{HTML}{d62728}

\begin{document}

\title{Neural Simulation-Based Inference of the Neutron Star Equation of State directly from Telescope Spectra}

\author{Len Brandes}
\email{len.brandes@tum.de}
\affiliation{Technical University of Munich,  TUM School of Natural Sciences,  Physics Department,  85747 Garching, Germany}
\author{Chirag Modi}
\affiliation{Center for Computational Astrophysics, Flatiron Institute, New York, NY, 11226, USA}
\affiliation{Center for Computational Mathematics, Flatiron Institute, New York, NY, 11226, USA}
\author{Aishik Ghosh}
\affiliation{Department of Physics and Astronomy, University of California, Irvine, CA 92697, USA}
\affiliation{Physics Division, Lawrence Berkeley National Laboratory, Berkeley, CA 94720, USA}
\author{Delaney Farrell}
\affiliation{Department of Physics, San Diego State University, San Diego, CA 92115, United States}
% \author{Daniel Whiteson}
% \affiliation{Department of Physics and Astronomy, University of California, Irvine, CA 92697, USA}
\author{Lee Lindblom}
\affiliation{Department of Physics,
University of California at San Diego, La Jolla, CA 92093, United States}
\author{Lukas Heinrich}
\affiliation{Technical University of Munich,  TUM School of Natural Sciences,  Physics Department,  85747 Garching, Germany}
\author{Andrew W. Steiner}
\affiliation{Department of Physics and Astronomy, University of Tennessee, Knoxville, Tennessee 37996, USA}
\affiliation{Physics Division, Oak Ridge National Laboratory, Oak Ridge, Tennessee 37831, USA}
\author{Fridolin Weber}
\affiliation{Department of Physics, San Diego State University, San Diego, CA 92115, United States}
\affiliation{Department of Physics,
University of California at San Diego, La Jolla, CA 92093, United States}
\author{Daniel Whiteson}
\affiliation{Department of Physics and Astronomy, University of California, Irvine, CA 92697, USA}

\begin{abstract}
Neutron stars provide a unique opportunity to study strongly interacting matter under extreme density conditions. %beyond the reach of laboratory experiments. 
The intricacies of matter inside neutron stars and their equation of state are not directly visible, but determine bulk properties, such as mass and radius, which affect the star's thermal X-ray emissions.  However, the telescope spectra of these emissions are also affected by the stellar distance, hydrogen column, and effective surface temperature, which are not always well-constrained. Uncertainties on these nuisance parameters must be accounted for when making a robust estimation of the equation of state. In this study, we develop a novel methodology that, for the first time, can infer the full posterior distribution of both the equation of state and nuisance parameters directly from telescope observations. 
This method relies on the use of neural likelihood estimation, in which normalizing flows use samples of simulated telescope data to learn the likelihood of the neutron star spectra as a function of these parameters, coupled with Hamiltonian Monte Carlo methods to efficiently sample from the corresponding posterior distribution. Our approach surpasses the accuracy of previous methods,  improves the interpretability of the results by providing access to the full posterior distribution, and naturally scales to a growing number of neutron star observations expected in the coming years.

\end{abstract}

\maketitle
\tableofcontents

%\vspace{.25in}

\section{Introduction}\label{sec:intro}

The study of neutron stars provides a window into the fundamental nature of matter under extreme conditions, a long-standing area of research in nuclear and astrophysics \cite{Baym2018}. The core of a neutron star contains matter at densities that far surpass those encountered in terrestrial experiments.
Under such conditions, matter could exist in various exotic states
\cite{Blaschke2018}, such as baryons in the form of hyperons and $\Delta$ isobars~\cite{Tolos2020,Li2018, Spinella2019,Malfatti2020,Sedrakian2023},  deconfined quarks~\cite{Alcock1986,Madsen1999,Weber2005,Orsaria2014},  color superconducting phases~\cite{Alford2001,Alford2008,Zdunik2013}, quarkyonic matter \cite{Buballa2005,McLerran2019}, or possibly meson condensates~\cite{Baym1973,Kaplan1986,Glendenning1999,Ellis1995,Ramos2001}. 
Understanding the physics of matter under the extreme conditions within a neutron star hinges crucially on unveiling the equation of state (EoS), which represents its internal composition through the intricate relationship between pressure and energy density \cite{Kojo2021a}. Knowledge of the EoS is also vital in determining the macroscopic properties of neutron stars, such as their mass and radius, via the relativistic stellar structure equations \cite{Tolman1939, Oppenheimer1939}.  

Observations of neutron stars have been the primary source for advancing our knowledge of the EoS of superdense matter. In the last two decades, there has been an increasing amount of high-quality data from X-ray and radio emissions, and gravitational waves \cite{Lattimer2014, Steiner2018, Ascenzi2024, Antoniadis2013, Arzoumanian2018, Fonseca2021, Riley2019, Miller2021, Miller2019, Miller2021, Abbott2019, Abbott2020}. In particular, spectroscopic measurements from the thermal surface emission of low-mass X-ray binaries in quiescence have provided constraints on the neutron star mass-radius relation. The constraints on the mass-radius relation from observations can be translated into constraints on the EoS of neutron star matter through the use of Bayesian inference techniques \cite{Raaijmakers2019, Legred2021, Brandes2023, Brandes2023a, Huth2022, Altiparmak2022, Lim2022, Marczenko2023, Annala2023, Han2023, Jiang2023, Essick2023, Mroczek2023b, Pang2023, Koehn2024}, with a recent interest in complementary methods based on machine learning (ML)
~\cite{Fujimoto2018,Fujimoto2020,Fujimoto2021,Ferreira2019,Morawski2020,Traversi2020,Krastev2021,Morawski2022,Ferreira2022,Thete2022,Soma2022,Soma2023,Krastev2023,Guo2023,Chatterjee2023,Zhou2023,Carvalho2023,Carvalho2024,Fujimoto2024,Zhou2024,Farrell2023a,Farrell2023b}. Beyond the star's mass and radius, emitted stellar spectra are also affected by physical quantities like the stellar distance, hydrogen column, and effective surface temperature. Uncertainties on these so-called nuisance parameters must be accounted for when analyzing stellar spectra and making a robust estimation of the equation of state. Traditionally, inference is done in two steps, first by extracting the mass and radius of a neutron star from an observed telescope spectrum, followed by EoS inference from a set of stellar masses and radii~\cite{Riley2018, Miller2019a}. This two-step procedure often requires additional assumptions, though recently it has been demonstrated that a single-step inference is possible while fully propagating uncertainties~\cite{Farrell2023a, Farrell2023b}.
While these advancements allow access to the posterior of the EoS parameters by marginalizing nuisance parameters, the computational expense involved has hindered access to the full posterior.

\begin{figure}[htb]
	\begin{center}
		\includegraphics[width=\textwidth,angle=-00]{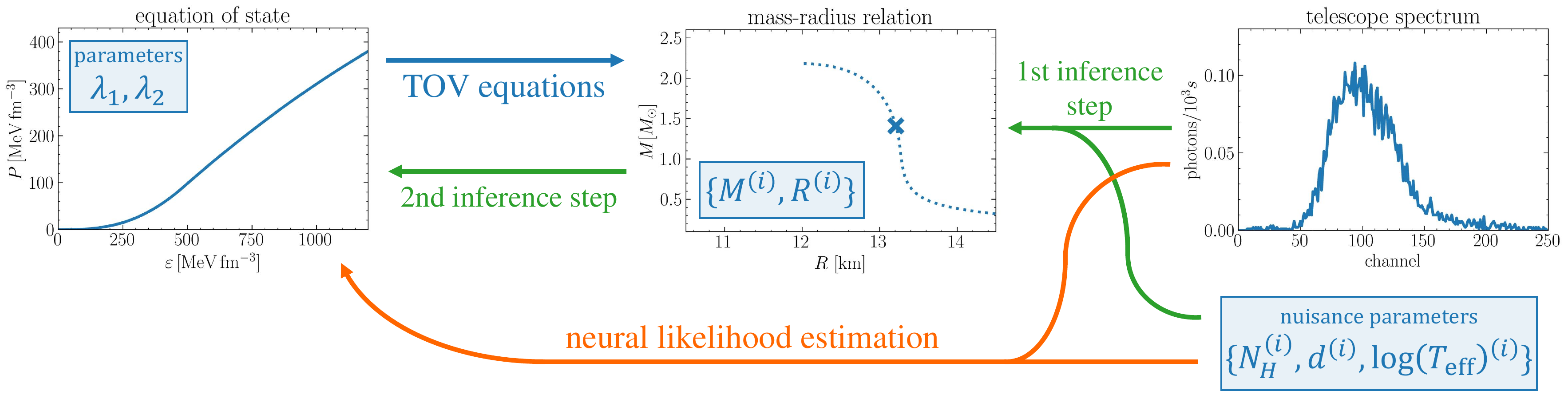}
        \caption{Traditional inference of EoS parameters (left) from telescope spectra (right) is done by first inferring intermediate mass-radius constraints (green arrows), involving additional implicit assumptions. In contrast, neural likelihood estimation allows for inference of the EoS directly
        from telescope spectra (orange arrow), robustly accounting for uncertainties.}
		\label{fig:illustration}
	\end{center}
\end{figure}

This paper introduces a novel method that provides access to the full posterior distribution of the equation of state and the nuisance parameters directly from telescope spectra. We employ a recently developed simulation-based inference technique known as neural likelihood estimation (NLE) \cite{Papamakarios2019} in which normalizing flows~\cite{Dinh2016,Papamakarios2017a} use samples of simulated X-ray telescope spectra to learn the likelihood of an observation as a function of the EoS and nuisance parameters. The direct inference of the neutron star EoS from telescope spectra via neural likelihood estimation in contrast to the traditional two-step inference is shown schematically in Fig.\,\,\ref{fig:illustration}.

%I (Aishik) like to have an illustration to explain the new idea within the first 2 pages.

Once trained, the likelihood evaluation is computationally inexpensive, allowing for the calculation of the posterior across the high-dimensional parameter space for inference, marginalization, profiling, or visualization.  Since neural networks are, by design, differentiable functions, we can efficiently sample the posterior using state-of-the-art Markov Chain Monte Carlo (MCMC) methods, such as Hamiltonian Monte Carlo (HMC). On simulated test spectra, this method outperforms previous approaches in terms of accuracy and precision of EoS estimation, while improving the interpretability of results with available posterior nuisance parameters. This approach also naturally scales to a growing number of neutron stars, as it does not require retraining to apply to larger datasets.

This paper is organized as follows: Section~\ref{sec:Background} reviews the connection between the equation of state of neutron stars and their observed X-ray spectra. This section also outlines the simulation process employed for generating samples of simulated telescope spectra. Sec.~\ref{sec:prviousWork} summarizes related previous work to infer the neutron star EoS. Sec.~\ref{sec:NLE} then describes our new simulation-based neural likelihood estimation (NLE) approach developed in this paper.
The performance of this method is quantitatively compared to previous approaches in Sec.~\ref{sec:results}, with a qualitative discussion on its merits in Sec.~\ref{sec:disandout}. Finally, we present our conclusions in  Sec.\,\,\ref{sec:conclusion}.

\section{Simulated neutron stars}
\label{sec:Background}

Samples of simulated neutron stars with varying values of EoS and nuisance parameters are prepared for training the normalizing flows and evaluation of their performance. Samples from Refs.~\cite{Farrell2023a,Farrell2023b} are used to facilitate direct comparison to previous methods.

\subsection{Equation of state }\label{subsec:eos}

Matter inside neutron stars can be described in terms of the EoS, which provides the thermodynamic relationship between pressure $P$ and energy density $\varepsilon$ within the star. 
To generate samples for the EoS, we begin with the relativistic mean field model GM1L~\cite{Typel2010} (see Appendix\,\,\ref{sec:RMF} for more details), represented by a second-order spectral expansion as described in Refs.~\cite{Lindblom2010, Lindblom2018}. Thus the full EoS can be described succinctly by the expansion coefficients, which we refer to as $\lambda_1$ and $\lambda_2$. We draw  EoS samples by uniformly sampling the coefficients in the intervals $\lambda_1 \in [4.75, 5.25]$ and $\lambda_2 \in [-2.05, -1.85]$.

The underlying physics captured in the star's EoS directly influences bulk properties like mass and radius through the relativistic stellar structure equations. We determine mass-radius relations for each EoS using the Tolman-Oppenheimer-Volkoff (TOV) equations, which determine the structure of non-rotating, spherically symmetric neutron stars in equilibrium \cite{Tolman1939, Oppenheimer1939}. Using the EoS variations described above, the TOV equations are numerically solved to produce mass-radius relations which are sampled to generate the mass $M$ and radius $R$ for $\sim$100 stars per EoS sample. 
For the sampling of stars we choose a log-uniform distribution of central enthalpies \cite{Lindblom1992} for the boundary condition to solve the stellar structure equations. This leads to a higher weighting of larger masses close to the maximum supported mass in the distribution of stars for each EoS.

\subsection{Modeling X-ray spectra with \xspec} \label{sec:xspec_spectra}

Traditional statistical methods infer macroscopic neutron star properties from the emitted X-ray spectra of quiescent low-mass X-ray binaries (LMXBs) by fitting the observed spectrum to well-motivated theoretical models~\cite{Miller2016, Miller2020, Heinke2006}. The open-source software \xspec contains many such models and is widely regarded as the state-of-the-art for spectral fitting~\cite{Arnaud1996} and can additionally be used in the generation of simulated spectra. The spectra used in this study were generated using the \texttt{NSATMOS} model in {\sc xspec}, a hydrogen atmospheric model with electron conduction and self-irradiation \cite{Heinke2006}. Beyond stellar mass and radius, this model also depends on three additional nuisance parameters, described in the next section.  The simulated spectra are subjected to the Chandra telescope response function corresponding to the instrument ACIS-S~\cite{Heinke2006,Bogdanov2016} and to Poisson noise corresponding to an observation time of $100\,$ks\footnote{The simulated spectra do not yield detectable photons at high energies. Therefore, we follow previous works \cite{Farrell2023a,Farrell2023b} to use only the first 250 bins of the telescope spectra.}.

\subsection{Nuisance parameters}

Each simulated X-ray spectrum from the \texttt{NSATMOS} model depends on the stellar mass, radius, and three additional nuisance parameters: the effective temperature of the surface, $T_{\mathrm{eff}}$, the distance to the star, $d$, and the hydrogen column, $N_H$, which parameterizes the reddening of the spectrum by the interstellar medium. Values for the nuisance parameters are sampled from ranges with distributions motivated by observation as described in Refs.~\cite{Steiner2018, Lattimer2014}; details  of each range are given in Tab.\,\ref{tab:ranges}.

\setlength{\extrarowheight}{4pt}
\begin{table}[htb]
    \centering
    \caption{Distributions and ranges of the equation of state parameters ($\lambda_1$ and $\lambda_2$) and nuisance parameters ($N_H$, $d$, and $\log (T_\mathrm{eff})$) used to generate the samples of simulated neutron stars. See text for parameter definitions.}    
    \label{tab:ranges}
    \begin{tabular}{llll}
        \hline \hline 
        %\multicolumn{4}{|c|}{Training samples} \\
        Parameter & & Distribution & Range \\ \hline
        EoS & $\lambda_1$ & uniform & [4.75, 5.25] \\
        & $\lambda_2$ & uniform & [-2.05, -1.85] \\ \hline
        nuisance & $d$ & uniform & [2.3, 12.3]$\,$kpc \\
        & $N_H$ & log uniform  \hspace*{5mm} & [0.01, 3.16]$\, 10^{21}\,$cm$^{-2}$ \\
        & $T_\mathrm{eff}$  \hspace*{5mm} & exponential & [1, 2]$10^6\,$K~\footnote{This corresponds to a uniform distribution of $\log(T_\mathrm{eff})$ in the range [6, 6.3].} \\ %\textcolor{red}{[$6 \times 10^5$, $2.4 \times 10^6$]$\,$K} 
        \hline \hline  
    \end{tabular}
    
\end{table}
\setlength{\extrarowheight}{4pt}

The values of nuisance parameters can be informed by independent observations, which can vary significantly from star to star and provide prior information. We consider three scenarios for nuisance parameter priors, referred to as `true', `tight', and `loose' \cite{Farrell2023a, Farrell2023b}. In the `true' scenario, the nuisance parameters are known exactly, while the `tight' and `loose' scenarios have narrow or wide Gaussian priors, respectively; see Tab.\,\,\ref{tab:priors} for a description of the prior widths for the three scenarios. 

\setlength{\extrarowheight}{4pt}
\begin{table}[htb]
    \centering
    \caption{ Prior distributions on nuisance parameters under three scenarios, `true', `tight', and `loose'. Shown are the widths of the Gaussian priors.}
    \label{tab:priors}
    \begin{tabular}{llll}
        \hline \hline 
%        &\multicolumn{3}{c}{Priors} \\
        Parameter &  true & tight & loose  \\ \hline
%        EoS & $\lambda_1$, $\lambda_2$ & \multicolumn{3}{l}{same as Tab.\,\ref{tab:priors}} \\ \hline
        %nuisance 
         $d$ & exact & 5\% & 20\% \\
         $N_H$ & exact & 30\% & 50\% \\
         $\log(T_\mathrm{eff})$  \hspace*{5mm} & exact \hspace*{1mm} & $\pm 0.1$  \hspace*{2mm} & $\pm 0.2$  \hspace*{2mm} \\
        \hline \hline  
    \end{tabular}
    
\end{table}
\setlength{\extrarowheight}{4pt}

\section{Previous work}
\label{sec:prviousWork}

The likelihood, $p(s|\lambda)$, of the spectra $s$ given the EoS parameters $\lambda$ is not analytically known. Many traditional methods that infer the EoS from telescope spectra approximate the unknown likelihood in a two-step method~\cite{Riley2018, Miller2019a}, first inferring posterior distributions $p(M,R|s)$ for the stellar mass and radius from an observed spectrum. Uncertainties on the nuisance parameters can produce non-trivial~\cite{Farrell2023a}, occasionally multimodal~\cite{Steiner2018} contours in the posterior probabilities of stellar mass and radius. Subsequently, these posterior distributions, approximated with a Kernel Density Estimator (KDE)\footnote{KDE corresponds to summing a Gaussian kernel at each posterior sample with a standard deviation equal to a bandwidth parameter $h$ \cite{Silverman1986}.}, are used as likelihoods in a second step to infer the EoS parameters \cite{Legred2021, Brandes2023, Brandes2023a, Huth2022, Altiparmak2022, Lim2022, Marczenko2023, Annala2023, Han2023, Jiang2023, Essick2023, Mroczek2023b, Pang2023, Koehn2024}. This is valid only if the $(M,R)$ priors used in the inference are sufficiently flat \cite{Riley2018,Raaijmakers2021}.   

Some previous machine learning (ML) approaches infer the EoS by focusing only on the second of the two steps, starting directly from the posterior probabilities of mass and radius. In Refs.~\cite{Fujimoto2018, Fujimoto2020, Fujimoto2021}, neural networks are used to perform regression of EoS parameters from a fixed number of stars described by their masses and radii, where the posterior probabilities are simplified as uncorrelated Gaussians. These neural networks need to be retrained when the number of measurements increases. 
Several other machine learning methods follow a similar approach with varying architectures~\cite{Ferreira2019, Morawski2020, Traversi2020, Krastev2021, Morawski2022, Ferreira2022, Thete2022, Soma2022, Soma2023, Krastev2023, Guo2023, Chatterjee2023, Zhou2023, Carvalho2023, Carvalho2024, Fujimoto2024}. 
The mass and radius standard deviations can be estimated directly from \xspec by varying the nuisance parameters~\cite{Farrell2023a}; a neural network approach that regresses EoS parameters from the resulting mass, radius and uncertainties is referred to below as `NN($M,R$ via {\sc xspec})'. 

However, deep neural networks are capable of analyzing high-dimensional inputs, allowing instead for regression of EoS parameters directly from a set of stellar spectra~\cite{Farrell2023a} and effectively removing the intermediate step in the two-step approach.
This method, referred to below as `NN(Spectra)', uses an uncertainty-aware and permutation-invariant neural network and captures the complex correlations between uncertainties, but only produces a point-estimate rather than the full posterior as a function of the EoS and nuisance parameters. Calculation of the full likelihood is intractable, but Ref.~\cite{Farrell2023b} showed that two neural networks can be used to replace the unavailable elements, granting access to the likelihood of the expected spectra given stellar mass, radius, and nuisance parameters, $p(s|M,R,\nu)$. While this method, referred to as `\surrEOS', succeeds in obtaining a posterior in EoS, it is computationally expensive to run. This makes it a prohibitively expensive method if access to the nuisance parameter posteriors is also desired. These three approaches are used for comparison in this work.

%%%%%%%%%%%%%%%%%%%%%%%%%%%%%%%%%%%%%%%%
\section{Bayesian inference with neural likelihood estimation}
\label{sec:NLE}

We introduce a single-step approach, which uses neural likelihood estimation to directly learn the likelihood of telescope spectra ($s$) as a function of the EoS parameters ($\lambda$) and the nuisance parameters ($\nu$) for a star, $p(s|\lambda,\nu)$. This extends the strategy in Ref.~\cite{Farrell2023b}, which learned $p(s|M,R,\nu)$, but avoids needing to integrate over the $M$-$R$ plane to achieve a connection to the EoS parameters, saving significant computational complexity. In addition, we apply a  
Hamiltonian Monte Carlo (HMC) method to efficiently draw samples from the posterior distribution, $p(\lambda,\nu|s)$.

%%%%%%%%%%%
\subsection{Neural likelihood estimation}

Neural likelihood estimation (NLE)~\cite{Papamakarios2019} is a type of 
simulation-based inference~\cite{Cranmer2020, Lueckmann2021} technique, successfully used for inference with gravitational waves~\cite{Delaunoy2020,Dax2021,Green2021}, particle physics~\cite{Cranmer2015,Brehmer2020,Ghosh2020} and cosmology~\cite{Alsing2019,Jeffrey2020,Hahn2023,Modi2023,Modi2023a} when the likelihood of the observed data are not analytically available but must be estimated from samples of simulated data.

In this approach a neural density estimator ($q_\Phi$, with parameters $\Phi$) approximates the likelihood:
\begin{align}
    q_\Phi(s|\lambda,\nu) \approx p(s|\lambda,\nu).
\end{align}

Normalizing flows (NF), powerful neural networks capable of modeling complicated probability distributions even in high dimensions~\cite{Papamakarios2017a}, are used as the density estimator. A brief introduction to NFs is provided in Appendix\,\,\ref{sec:NormalizingFLow}. Once trained, NFs can easily generate new samples as well as estimate the likelihood of a given sample.  Specifically, a Masked Autoregressive Flow (MAF)~\cite{Papamakarios2017b} is used for the density estimator $q_\Phi$. To fit the parameters $\Phi$ such that the learned distribution $q_\Phi$ approximates the target likelihood distribution $p$, we minimize the Kullback-Leibler divergence ($D_\text{KL}$) between the two distributions, which is equivalent to maximizing the approximate log-probability of the samples generated from the likelihood distribution: 
\begin{align}
    \arg\min_{\Phi}  D_\text{KL} \left(p(s|\lambda,\nu) \big|\big|q_\Phi(s|\lambda,\nu) \right) &=  \arg\max_{\Phi}  \sum_{s_i\sim p(s|\lambda_i, \nu_i)}  \log q_\Phi(s_i|\lambda_i,\nu_i) ~.
\end{align}

A derivation of this equivalence is given in Appendix\,\,\ref{sec:NormalizingFLow}.
With samples $s_i$ drawn from the likelihood via simulation, as described above,
\begin{align}
    s_i \sim p(s|\lambda_i,\nu_i) ~,
\end{align}

\noindent
the \verb|sbi| package~\cite{TejeroCantero2020} is used to train the MAF as a neural likelihood estimator. To make our analysis robust to stochasticities in training the neural network, we perform a hyperparameter search by training 100 different MAFs with different architectures. After training, we use an ensemble average~\cite{Hermans2021} over the $N=5$ best-performing density estimators such that the log-likelihood of any given telescope spectrum $s_0$ is 
\begin{align}
    \log p(s_0| \lambda, \nu) \approx \frac{1}{N} \sum_j \log q_{\Phi_j}(s_0| \lambda, \nu) ~.
\end{align}

%%%%%%%%%%%%%%%
\subsection{Posterior sampling with Hamiltonian Monte Carlo}

The posterior distribution can be built from the estimated likelihood and prior distributions:
\begin{align}
    p(\lambda, \nu|s) &\propto p(s|\lambda, \nu) p(\lambda) p(\nu) ~.
\end{align}
The prior distribution on the nuisance parameters, $p(\nu)$, is given in Tab.\,\ref{tab:priors}, while the prior on the EoS parameters, $p(\lambda)$, is taken to be uniform in the intervals $\lambda_1 \in [4.75, 5.25]$ and $\lambda_2 \in [-2.05, -1.85]$, the same as the distribution of training samples described in Tab.\,\ref{tab:ranges}. 

To draw samples from the posterior distribution, sampling algorithms like Markov Chain Monte Carlo (MCMC) or nested sampling can be employed. Given that the normalizing flows, priors and, consequently, the full posterior distribution are differentiable, we draw samples from the posterior using Hamiltonian Monte Carlo (HMC) \cite{Neal2011,Betancourt2017} sampling, which can use the gradient information and scales much more efficiently to high dimensional parameter spaces. For a brief introduction to HMC and our implementation details, see Appendix\,\,\ref{sec:HMC}.

Unlike standard approaches, our methodology allows for the simultaneous inference of EoS parameters $\lambda$ and nuisance parameters $\nu$, which minimizes assumptions made about these parameters and therefore makes the approach more robust. Additionally, any supplemental information on these parameters coming from other observations can be naturally included in the analysis through the prior distribution $p(\nu)$ without retraining the neural density estimators.

\subsection{Scaling to multiple observations}

The estimation of the per-star likelihood as a function of the EoS parameters, $p(s|\lambda, \nu)$, makes the calculation of the joint likelihood for $J$ stars straightforward:

\begin{align}
    p(s_{1...J}|\lambda, \nu_{1...J}) = \prod_j  p(s_j|\lambda,\nu_j)  ~.
    \label{eq:MultipleSpectra}
\end{align}
Each star has a specific set of nuisance parameters, $\nu_j$, such that the posterior is
\begin{align}
    p(\lambda, \nu_{1...J}|s_{1...J}) &\propto  
    \left( \prod_j\, p(s_j|\lambda,\nu_j) p(\nu_j) \right) p(\lambda) ~.
\end{align}

 Scaling to multiple observations is straightforward for several reasons. Estimation of the likelihood in the EoS parameters rather than $M$ and $R$, means that no additional integration over the $M$-$R$ plane is required. In addition, the likelihood itself is estimated, rather than posteriors which cannot be trivially combined~\cite{Papamakarios2018,Greenberg2019,Durkan2020,Hermans2020}. Learning the likelihood conditioned on the nuisance parameters for each star rather than implicitly marginalizing over them allows for the joint likelihood to be simply a product of the individual stellar likelihoods.

A consequence of these choices is that we infer not only the equation of state parameters but also the nuisance parameters corresponding to every star. The stellar nuisance parameter priors are independent and can encode prior information for each star separately, for maximum flexibility and robustness. 
While this increases the dimensionality of the problem, 
the availability of gradients of the posterior distribution enables the use of powerful algorithms like HMC, ensuring that the inference remains computationally tractable.

%%%%%%%%%%%%%%%%%%%
\section{Results}
\label{sec:results}

Inference of the full posterior distribution of neutron star EoS and nuisance parameters directly from telescope spectra is now feasible. Following Refs.~\cite{Farrell2023a,Farrell2023b}, we apply the method to a set of ten simulated neutron stars for a given point in EoS space, each with three independent nuisance parameters defined above. The complete parameter space comprises two parameters of interest and thirty nuisance parameters. 
Results are shown for a single representative EoS point, we then demonstrate the scaling of the NLE approach to more observations and finally present a comparison with previous approaches by evaluating the average performance over 100 randomly sampled test points in EoS space. 

\subsection{Example posterior distribution}

\begin{figure}[htb]
	\begin{center}
		\includegraphics[height=150mm,angle=-00]{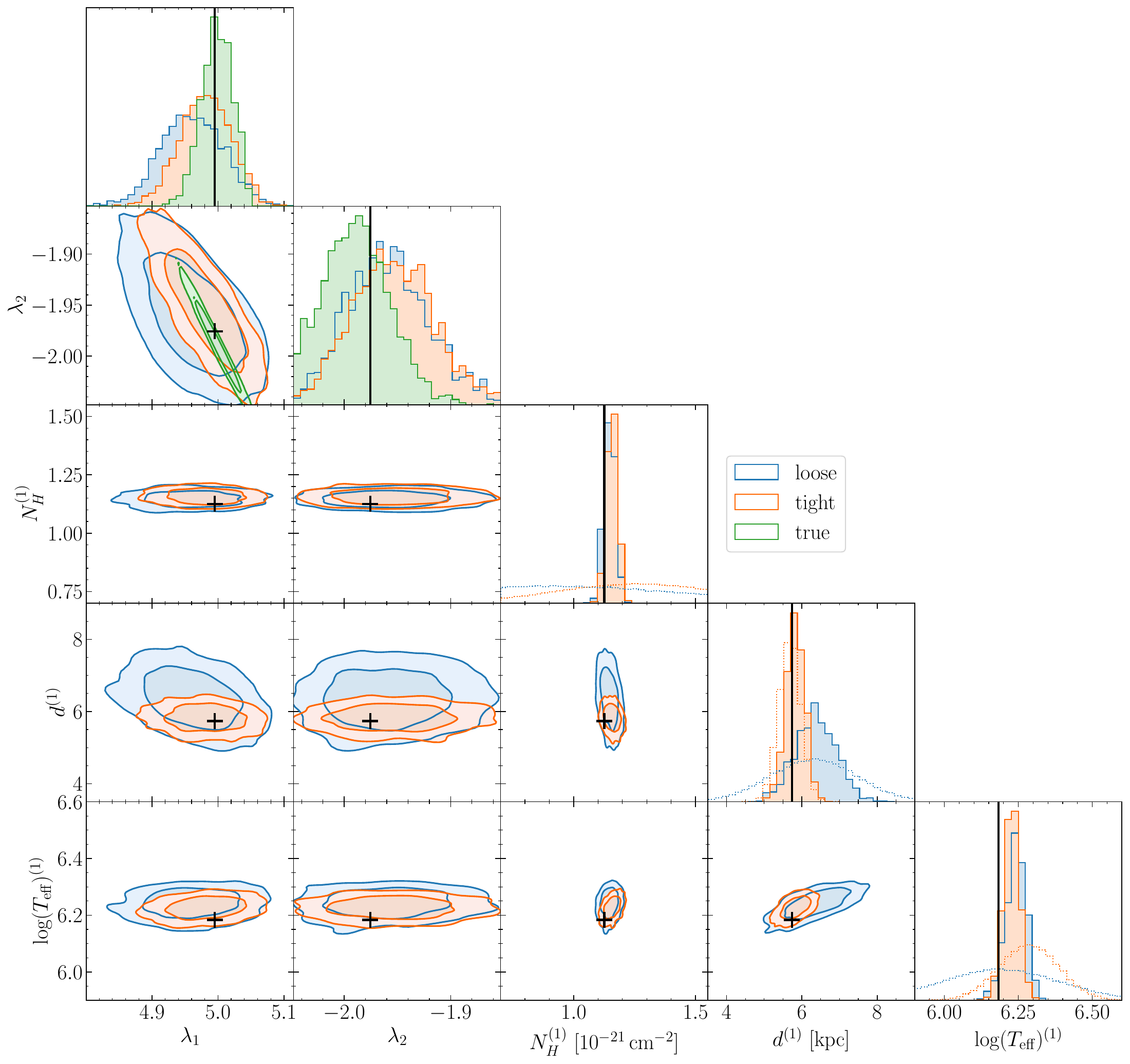}
		\caption{Corner plot depicting the posterior distribution of the parameters $\lambda_1$ and $\lambda_2$  of one example EoS as well as the first 3 (of 30) nuisance parameters $N_{H}^{(1)}$, $d^{(1)}$ and $\log (T_\mathrm{eff})^{(1)}$. The posterior is computed based on the simulated spectra of 10 stars with the nuisance parameters known exactly in the true scenario (green), and known with the uncertainties in Tab.\,\ref{tab:priors} in the tight (orange) and loose (blue) scenarios. The ground-truth parameter values are depicted as black crosses/lines. The marginal posterior distributions of the nuisance parameters are compared to the respective priors (dotted) of the tight and loose scenarios.}
		\label{fig:corner_plot}
	\end{center}
\end{figure}

An example of the posterior distributions marginalized to one and two dimensions is shown in the corner plot in Fig.~\ref{fig:corner_plot}. The figure depicts both EoS parameters, $\lambda_1$ and $\lambda_2$, and the nuisance parameters for the first neutron star, $N_{H}^{(1)}$, $d^{(1)}$ and $\log (T_\mathrm{eff})^{(1)}$. While the marginalized posteriors of the nuisance parameters for the other nine stars are also available, they are not shown here. In all three nuisance parameter scenarios from Tab.\,\ref{tab:priors}, the EoS parameters are strongly correlated, similar to the log-likelihoods computed in Ref.~\cite{Farrell2023b}. The marginal posterior distribution of $\lambda_1$ is relatively tight compared to its prior range, while for the second parameter $\lambda_2$ it is not as well constrained compared to the parameter's prior range. 

As expected, in the true scenario where the nuisance parameters are exactly known, the marginal posterior distributions are sharply centered around the ground-truth values. In the tight scenario, the uncertainty in the nuisance parameter distributions leads to wider distributions for the EoS parameters. This is further pronounced for the loose case, where less prior information on the nuisance parameters is available. Fig.\,\,\ref{fig:corner_plot} illustrates that the hydrogen column $N_H$ as well as the logarithm of the effective surface temperature $\log (T_\mathrm{eff})$ can be significantly constrained from the spectrum data compared to their prior ranges. In the tight scenario, the marginal posterior for the distance $d$ is almost indistinguishable from the prior, indicating that the telescope spectra do not contribute any more information for this parameter over the tight priors. However, in the loose case, the marginal posterior distribution of $d$ becomes tighter than the loose prior, which implies that we can indeed extract information about the distance of a neutron star from its X-ray spectrum. 

\begin{figure}[htb]
	\begin{center}
		\includegraphics[height=55mm,angle=-00]{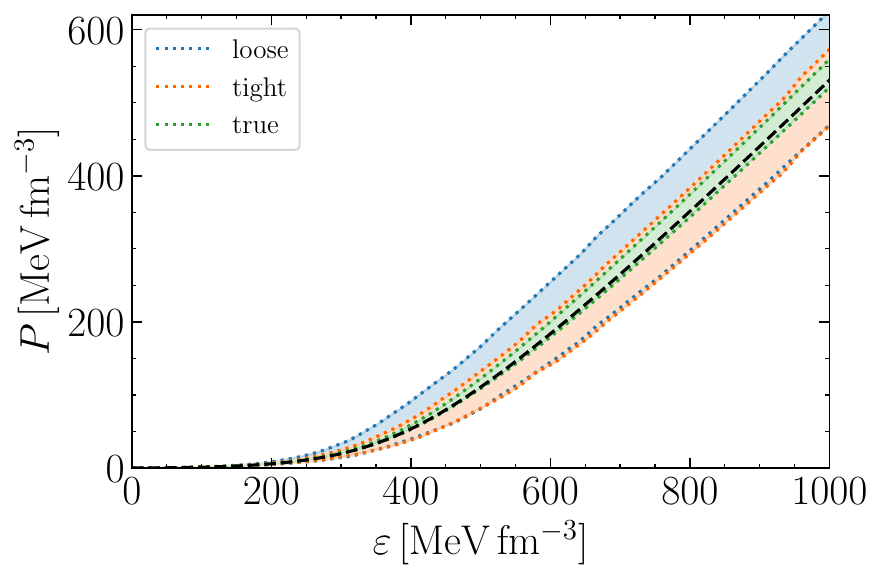}
        \includegraphics[height=55mm,angle=-00]{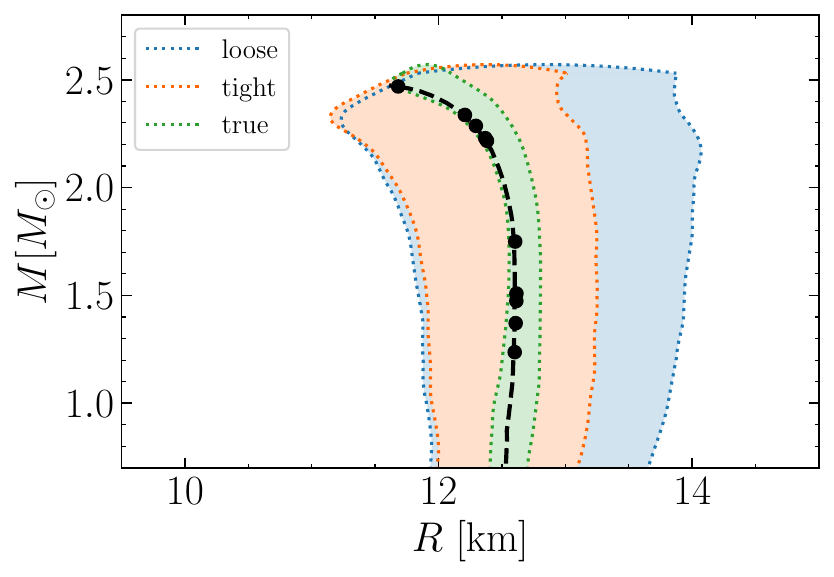}
		\caption{Posterior 95\% (highest density) credible bands for the pressure as a function of energy density and the radius as a function of mass for the three (true, tight, loose) scenarios. Similar to Fig.\,\,\ref{fig:corner_plot}, the posterior is derived based on the simulated spectra of 10 stars. The ground-truth value for the equation of state and the corresponding mass-radius relation is depicted as a dashed black line. Black dots indicate ground-truth mass-radius values of the 10 simulated stars.}
		\label{fig:credible_bands}
	\end{center}
\end{figure}

\begin{figure}[htb]
	\begin{center}
		\includegraphics[height=55mm,angle=-00]{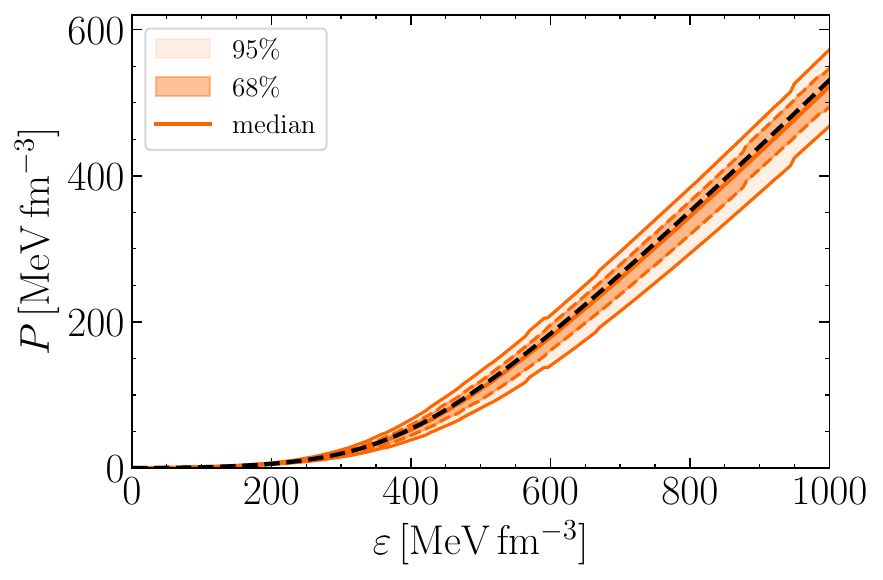}
        \includegraphics[height=55mm,angle=-00]{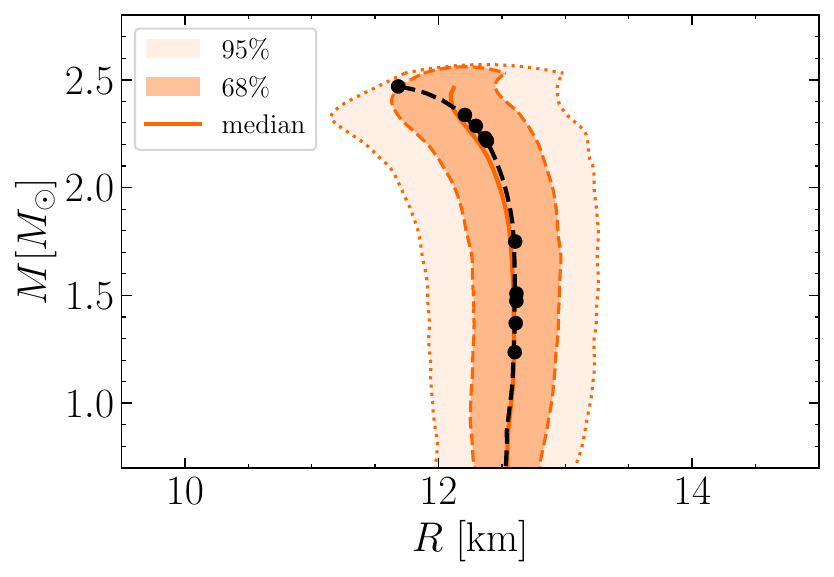}
		\caption{Similar to Fig.\,\,\ref{fig:credible_bands}, but only for the \textit{tight} scenario the median and the 68\% and 95\% posterior credible bands are shown for the pressure as a function of energy density and the radius as a function of mass. The ground-truth value for the equation of state and the corresponding mass-radius relation is depicted as a dashed black line. Black dots indicate the mass-radius values of the 10 simulated stars.}
		\label{fig:credible_bands_tight}
	\end{center}
\end{figure}

We can transform the posterior distribution for the EoS parameters $\lambda_1$ and $\lambda_2$ into 95\% (highest density) posterior credible bands for the pressure as a function of energy density as depicted in Fig.\,\,\ref{fig:credible_bands}. %To calculate the credible bands, each EoS is employed up to its maximum central energy density corresponding to the maximum mass endpoint of its mass-radius relation [TODO]. 
As before, the constraints are much tighter in the true scenario and become increasingly broader in the tight and loose cases. By solving the TOV equations, we can translate the EoS into mass-radius constraints. The 95\% posterior credible bands for the radius as a function of mass are depicted in Fig.\,\,\ref{fig:credible_bands}. The credible bands terminate at the 95\% upper limits of the maximum mass. Focusing only on the tight case in Fig.\,\,\ref{fig:credible_bands_tight}, there is a very close agreement of the inferred median for $P(\varepsilon)$ and $R(M)$ to the ground-truth values.  %[TODO discuss the correlation between EoS and nuisance parameters?]
Note that for this particular example, the mass of one of the simulated stars is very close to the respective maximum supported mass such that a good reconstruction even of the high-density part for this particular EoS is possible. In other cases, the reconstruction might be worse. 

\subsection{Increasing the number of observations}
\label{sec:observationnumber}

\begin{figure}[htb]
	\begin{center}
		\includegraphics[height=85mm,angle=-00]{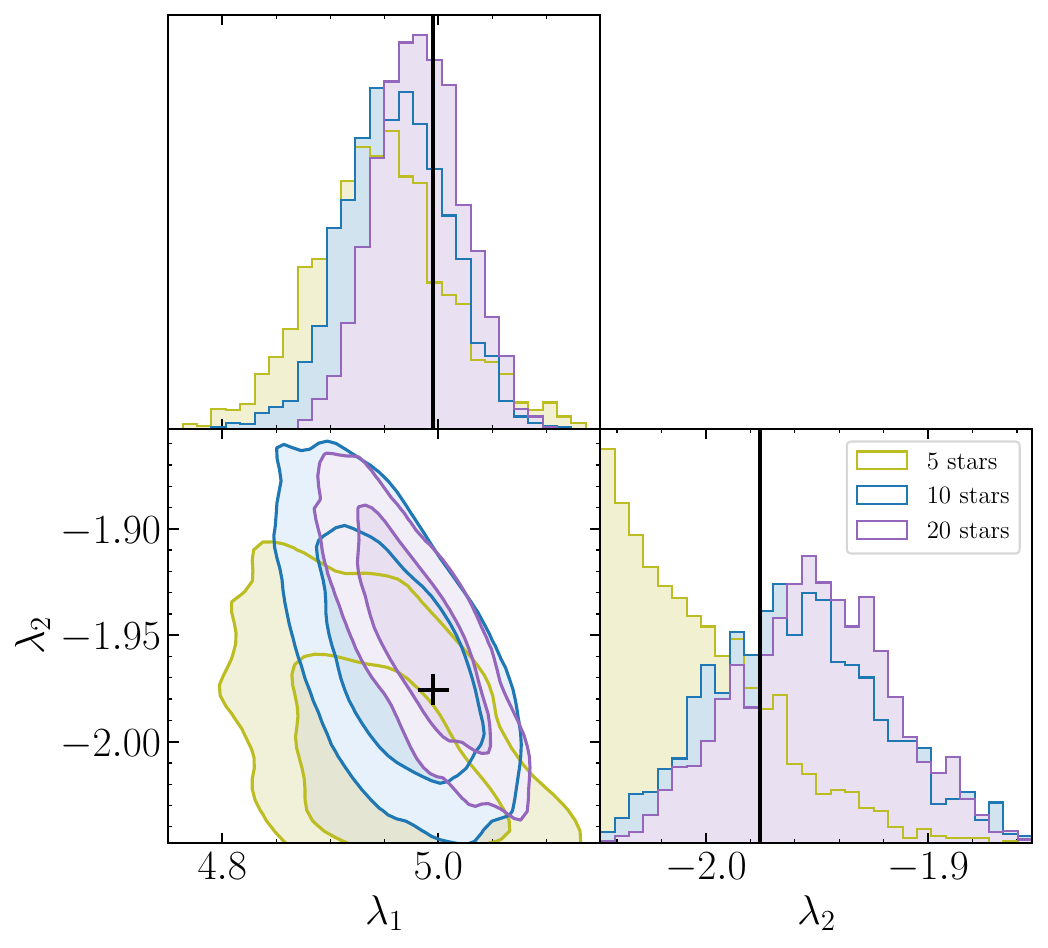}
		\caption{Corner plot depicting the posterior distribution of the parameters $\lambda_1$ and $\lambda_2$ of one example EoS. The posterior is computed based on the simulated spectra of 5 (olive), 10 (blue), or 20 (purple) stars with the nuisance parameters known with the uncertainties in Tab.\,\ref{tab:priors} of the \textit{loose} scenario. The ground-truth parameter values are depicted as black crosses/lines. }
		\label{fig:num_obs}
	\end{center}
\end{figure}

With an anticipated surge in the number of available neutron star observations in the upcoming years, the inference method must be able to scale to a large set of data. In our approach, normalizing flows approximate the likelihood $p(s|\lambda,\nu)$ per observed star. These likelihoods are then combined to obtain the total likelihood for a set of neutron stars, see Eq.\,(\ref{eq:MultipleSpectra}). 
Consequently, the method does not need to assume a fixed number of observed neutron stars, nor a particular ordering of the stars, and works out-of-the-box for a growing set of observed neutron stars without the need for retraining any networks. To demonstrate this, we present the marginal posterior distributions for one example EoS in Fig.\,\,\ref{fig:num_obs}, for 5, 10, and 20 observed neutron stars, and with loose uncertainties on the nuisance parameters.

The figure illustrates that the increase of available spectra significantly refines the inference of the EoS parameters. Notably, the transition from 5 to 10 observed spectra has a substantial impact on the posterior constraints, reducing the standard deviation of $\lambda_1$ by 23\% from 0.061 to 0.047, while further increasing the number of measurements to 20 shows a comparatively smaller reduction in the standard deviation by only 6.4\% to 0.044 for the given example. 
For $\lambda_2$ the increase in accuracy from 10 to 20 observations is even smaller. It is worth noting that in the numerical implementation of Hamiltonian Monte Carlo for posterior sampling, the computation time is predominantly consumed by the evaluation of the likelihood and its gradient. While the availability of more observations increases the per-iteration computational time in sampling the posterior, it also speeds up the convergence of the algorithm. Furthermore, HMC algorithms can easily be scaled to thousands of dimensions, hence we do not anticipate the dimensionality to be a limiting factor in the scaling of our approach. 

\subsection{Average performance on test set}
\label{sec:performance}

\begin{figure}[htb]
	\begin{center}
		\includegraphics[height=55mm,angle=-00]{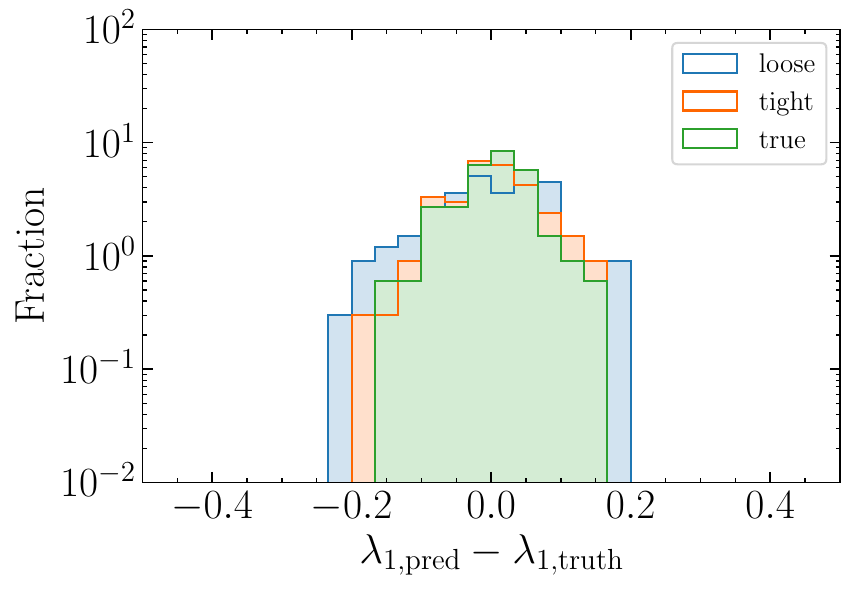}
        \includegraphics[height=55mm,angle=-00]{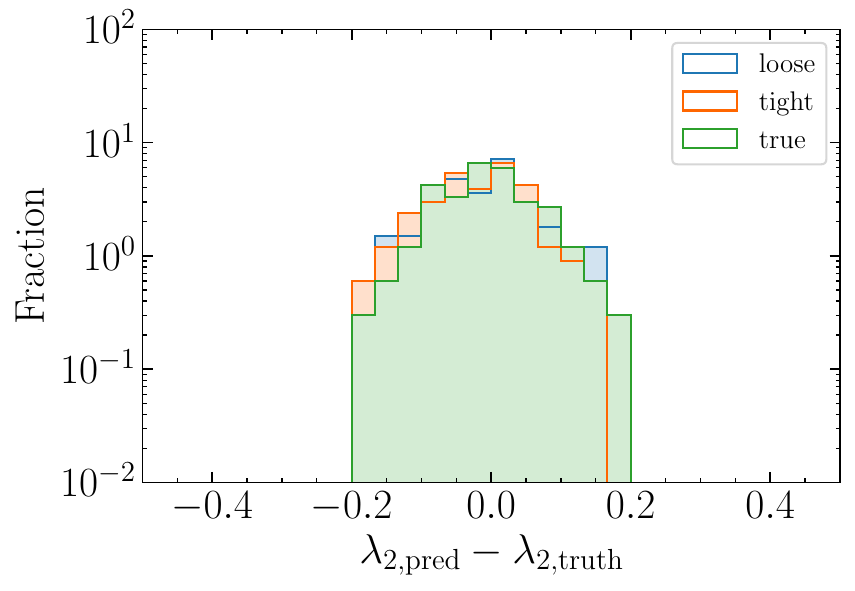}
		\caption{Distribution of the predicted maximum-a-posteriori estimates versus the ground-truth values for the two equation of state parameters $\lambda_1$ and $\lambda_2$. In the true scenario, the nuisance parameters are fixed to their exact values; in the tight and loose cases, they are drawn from the narrow or wide priors in Tab.\,\ref{tab:priors}.}
		\label{fig:diffs}
	\end{center}
\end{figure}

After discussing one example EoS, now we turn to the average performance of NLE with a test set of simulated data from 100 different equations of states. To compare the average performance to the previous ML approaches that infer the neutron star EoS directly from telescope spectra described in Sec.\,\,\ref{sec:prviousWork}, we use the same accuracy measure as Refs. \cite{Farrell2023a,Farrell2023b}. For each EoS in the test set, we simulate 10 spectra with random nuisance parameters. Based on the spectra and the prior nuisance parameter information, we then sample the posterior using the methodology outlined in Sec.\,\,\ref{sec:NLE}. From the marginal posterior distributions, similar to the example in Fig.\,\,\ref{fig:corner_plot}, we determine the maximum-a-posteriori estimates (MAP)\footnote{Note that this is the maximum-a-posteriori estimate of the marginal and not of the full posterior.} for the two EoS parameters $(\lambda_{1,\mathrm{pred}}, \lambda_{2,\mathrm{pred}})$ and compare them to the ground-truth values $(\lambda_{1,\mathrm{truth}}, \lambda_{2,\mathrm{truth}})$. The distributions of the differences between the marginal MAP estimates and the ground-truth values, $(\lambda_{1,\mathrm{pred}} - \lambda_{1,\mathrm{truth}},  \lambda_{2,\mathrm{pred}} - \lambda_{2,\mathrm{truth}})$, on the test data set are depicted in Fig.\,\,\ref{fig:diffs}. As before, the equation of state parameters can be maximally constrained in the true nuisance parameter scenario. In all three scenarios, the distributions are centered around zero, indicating that there is no systematic bias in the posterior prediction. Compared to the previous ML analyses in Refs.\,\,\cite{Farrell2023a,Farrell2023b}, the distributions of $\lambda_1$ are much narrower in the tight and loose scenarios (see for example Fig.\,\,8 in \cite{Farrell2023b}). 

\begin{figure}[htb]
	\begin{center}
		\includegraphics[width=\textwidth,angle=-00]{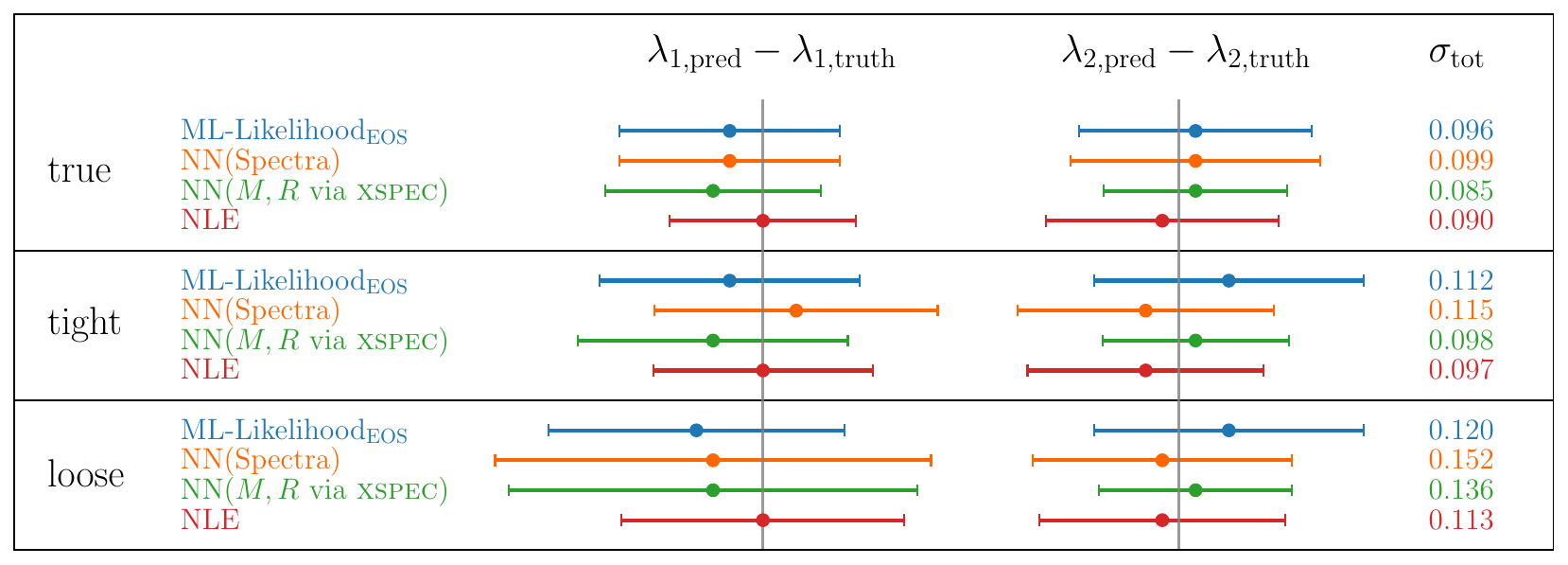}
        \caption{Illustrated mean and standard deviation of the difference between the predicted maximum-a-posteriori values to the ground-truth values for the three different scenarios from Tab.\,\ref{tab:eos}.}
		\label{fig:diffs_mu_sigma}
	\end{center}
\end{figure}

To quantitatively compare these distributions to the previous ML analyses, we compute the mean $\mu$ and standard deviation $\sigma$ of the distribution of differences in Fig.\,\,\ref{fig:diffs}. 
We can combine both standard deviations into 
\begin{align}
    \sigma_\mathrm{tot} = \sqrt{\sigma_{\lambda_1}^2 + \sigma_{\lambda_2}^2} ~. 
    \label{eq:sigmatot}
\end{align}
The resulting values are listed in Tab.\,\ref{tab:eos} and illustrated in Fig.\,\,\ref{fig:diffs_mu_sigma}. The mean $\mu$ of the difference between the posterior prediction of $\lambda_1$ and its ground-truth value is much smaller in the NLE approach compared to previous approaches in all three prior distributions considered in Tab.\,\ref{tab:priors}. 
In the true case, the NLE approach performs better than \surrEOS\hspace*{0mm} from Ref.~\cite{Farrell2023b} and NN(Spectra) from Ref.~\cite{Farrell2023a} (see Sec.\,\,\ref{sec:prviousWork} for more details about these methods). For realistic scenarios of nuisance parameters (loose and tight), NLE outperforms all other methods as quantified by $\sigma_\mathrm{tot}$, while for the true scenario, it outperforms all methods besides NN($M,R$ via {\sc xspec}), which uses \xspec itself for part of the inference and involves simplifying assumptions about the mass-radius uncertainties.
\begin{table}[htb]
    \centering
    \caption{Average accuracy for the prediction of neutron star EoS parameters $\lambda_1$ and $\lambda_2$. Shown are the means ($\mu$) and standard deviations ($\sigma$) of the distributions in Fig.~\ref{fig:diffs}, i.e., of the differences between the predicted maximum-a-posteriori and ground-truth values. Both standard deviations are combined to $\sigma_\text{tot}$ according to Eq.\,(\ref{eq:sigmatot}).
    The neural likelihood estimation (NLE) approach is compared to three previous approaches; neural networks that regress the EoS parameters from the spectra (NN(Spectra)) and from $M,R$ estimates by {\sc xspec} (NN($M,R$ via {\sc xspec})), both from Ref.~\cite{Farrell2023a}, as well as an approach using an approximate likelihood that incorporates two neural networks, \surrEOS,  from \cite{Farrell2023b}. In the true scenario, the nuisance parameters are fixed to their exact values; in the tight and loose cases, they are drawn from the narrow or wide priors in Tab.\,\ref{tab:priors}.}
    \label{tab:eos}
    \begin{tabular}{llrrrrrrc}
             \hline \hline
         & &\multicolumn{2}{c}{$\lambda_{1,\mathrm{pred}} - \lambda_{1,\mathrm{truth}}$} & \hspace*{10mm}  &\multicolumn{2}{c}{$\lambda_{2,\mathrm{pred}} - \lambda_{2,\mathrm{truth}}$} &&  Combined\\
         %\cline{3-4}\cline{6-7}\\
         $p(\nu)$ & Method & \multicolumn{1}{c}{$\mu$} & \multicolumn{1}{c}{\hspace*{5mm}$\sigma$} & & \multicolumn{1}{c}{$\mu$} & \multicolumn{1}{c}{\hspace*{5mm} $\sigma$} & \hspace*{8mm} &  $\sigma_\mathrm{tot}$\\
                  \hline 
        true \hspace*{10mm} & \surrEOS   \hspace*{10mm}       & -0.02 & 0.066  && \textbf{0.01}  & 0.070   &&  0.096 \\
        & NN(Spectra)          & -0.02 & 0.066  && \textbf{0.01}  & 0.075   && 0.099  \\
        & NN($M,R$ via {\sc xspec}) & -0.03 & 0.065  && \textbf{0.01}  & \textbf{0.055}   && \textbf{0.085} \\
        & \textbf{NLE} & \textbf{0.00} & \textbf{0.056} && \textbf{-0.01} &  0.070 &&  0.090 \\ \hline
        tight & \surrEOS         & -0.02 & 0.078  && 0.03  & 0.081   && 0.112 \\
        & NN(Spectra)          & 0.02  & 0.085  && -0.02 & 0.077   && 0.115  \\
        & NN($M,R$ via {\sc xspec}) & -0.03 & 0.081  && \textbf{0.01}  & \textbf{0.056}   && 0.098 \\ 
        & \textbf{NLE} & \textbf{0.00} & \textbf{0.066} && -0.02 & 0.071 && \textbf{0.097} \\ \hline
        loose & \surrEOS         & -0.04 & 0.089  && 0.03  & 0.081   && 0.120  \\
        & NN(Spectra)          & -0.03 & 0.131  && \textbf{-0.01} & 0.078   && 0.152 \\
        & NN($M,R$ via {\sc xspec}) & -0.03 & 0.123  && \textbf{0.01}  & \textbf{0.058}   && 0.136  \\
        & \textbf{NLE} & \textbf{0.00} & \textbf{0.085} && \textbf{-0.01} & 0.074 &&  \textbf{0.113} \\
         \hline \hline
    \end{tabular}
\end{table}
Interestingly, from Tab.\,\ref{tab:eos} it becomes clear that the NLE approach is better than all other approaches to constrain the first EoS parameter $\lambda_1$, while for $\lambda_2$ the accuracy of the \xspec based two-step approach, with its simplifying assumptions regarding the impact of nuisance parameters, is always much better than all other approaches. Note that the central enthalpies used to solve the stellar structure equations are sampled from log-uniform intervals, such that the masses near the maximum supported mass are weighted more heavily for each EoS. Consequently, for each EoS in the test set, the largest of the ten randomly selected masses is close to the respective maximum supported mass. This is also true for the previous approaches with which we compare our results.

With our neural likelihood estimation approach, we can now, for the first time, infer the posterior for nuisance parameters. The means and standard deviations of the differences between the marginal MAP estimates and the ground-truth values on the test data are listed in Tab.\,\,\ref{tab:np}.  
While we obtain excellent constraints on the hydrogen fraction $N_H$ and the effective surface temperature $\log (T_\mathrm{eff})$, our ability to constrain the distance $d$ from the spectra is limited. When going from the tight to the loose scenario, the inference on the distance estimates suffers the most, indicating that these constraints are primarily driven by the prior in the tight case, as seen in Fig.\,\,\ref{fig:corner_plot}. However, this is not as limiting as it might seem because it is much easier to obtain precise constraints on the distance of a neutron star from external measurements than it is for the other nuisance parameters.

\begin{table}[htb]
    \centering
    \caption{Similar to Tab.\,\ref{tab:eos}, but with mean and standard deviation of the predicted maximum-a-posteriori estimate versus the ground-truth value for the three nuisance parameters.}
    \begin{tabular}{llrcrrcrrc}
        \hline \hline
         & &\multicolumn{2}{c}{$N_{H,\mathrm{pred}} - N_{H,\mathrm{truth}}$} & \hspace*{10mm} & \multicolumn{2}{c}{$d_\mathrm{pred} - d_\mathrm{truth}$} & \hspace*{10mm}& \multicolumn{2}{c}{$\log (T_\mathrm{eff})_\mathrm{pred} - \log( T_\mathrm{eff})_\mathrm{truth}$} \\
         %\cline{3-4}\cline{6-7}\cline{9-10}\\
         Method \hspace*{10mm} & $p(\nu)$ \hspace*{10mm} &  \multicolumn{1}{c}{$\mu$} & \multicolumn{1}{c}{$\sigma$} & & \multicolumn{1}{c}{$\mu$} & \multicolumn{1}{c}{$\sigma$} & & \multicolumn{1}{c}{$\mu$} & \multicolumn{1}{c}{$\sigma$}\\ \hline 
        NLE & tight & 0.00 & 0.063 && -0.06 \hspace*{4mm} & 0.419 && 0.00 & 0.034 \\ 
         & loose & 0.01 & 0.070 && -0.26 \hspace*{4mm}  & 1.149 &&  -0.01 & 0.047 \\
         \hline \hline
    \end{tabular}
    \label{tab:np}
\end{table}

\section{Discussion \& outlook}

\label{sec:disandout}

In addition to the performance gains seen in the preceding sections, the NLE+HMC approach has several advantages relative to previous work.

\begin{enumerate}

\item[(i)] The single-step estimation of the likelihood in terms of the EoS avoids collapsing the information into the mass-radius plane as an intermediate step. EoS-dependent quantities like temperature inhomogeneities~\cite{Elshamouty2016} can impact a star’s spectrum but are not captured by the mass and radius. Uncertainties from nuisance parameters can also be more accurately propagated by using the full high-dimensional data. In addition, this avoids assuming the mass-radius posteriors can be used as a likelihood, making simplifying approximations regarding the nature of the intermediate likelihood, or integrating over the $M$-$R$ plane.

\item[(ii)] Neural likelihood estimation allows for amortization; after training the neural density estimators \textit{once}, the inclusion of additional observations is straightforward, see Sec.\,\,\ref{sec:observationnumber}. In addition, extending to additional stars is inexpensive relative to other methods, which require integrating over estimated mass-radius posteriors to construct likelihoods~\cite{Brandes2023a,Koehn2024}, such as with Kernel Density Estimation techniques.

\item[(iii)] Learning the likelihood instead of the posterior allows combination with likelihoods from other data  \cite{Lange2018}, e.g., constraints from low-energy nuclear theory at small densities \cite{Drischler2022,Keller2023}, perturbative QCD at high densities \cite{Gorda2021,Komoltsev2022}, mass measurements from Shapiro time delays \cite{Antoniadis2013,Arzoumanian2018,Fonseca2021}, mass-radius constraints from analyses of the NICER telescope \cite{Riley2019,Miller2021,Miller2019,Miller2021} or gravitational wave signals from binary neutron star mergers \cite{Abbott2019,Abbott2020}. If these likelihoods are computed by traditional methods, they may not be differentiable, which precludes the use of HMC sampling methods, however it may be worthwhile extending simulation-based inference techniques also to these cases.

\item[(iv)] Application of HMC allows for robust exploration of the high-dimensional space of nuisance parameters, which can be of interest in other astrophysical studies.

\end{enumerate}

Improvements to the method could be pursued in several directions.  To prevent degradation of performance near the borders of the training data, additional simulated samples could be included beyond the current borders.   Another potential improvement may be to 
additionally condition the inference on the masses $M$ of the neutron stars, approximating $p(s|\lambda,\nu,M)$ rather than $p(s|\lambda,\nu)$, to further enhance the interpretability and accuracy of the method. In this way, we could infer the neutron star masses and, by solving the TOV equations, radii simultaneously with the EoS.  Moreover, to combine the information from different observational instruments for the same neutron star, it becomes essential to compute the likelihood as a function of the neutron star mass. Such situations arise regularly, see e.g. \cite{Steiner2018,Klochkov2014,Doroshenko2022,Riley2021,Posselt2022}. Preliminary results show, however, that the mass distributions can become multimodal, as also seen in \cite{Steiner2018}, which makes sampling the posterior difficult. Future work may extend our method to effectively include the mass as a nuisance parameter.
In addition, one could explore alternative parameterizations of the EoS which can describe more complicated phase structures possibly realized inside neutron stars like crossovers of first-order phase transitions, such as piecewise polytropes \cite{Greif2019}, segment-wise linear interpolations of the speed of sound \cite{Annala2020}, non-parametric representations based on a Gaussian process \cite{Landry2020,Legred2021} or a neural network~\cite{Han2021}. In that way, we could constrain the unknown phase structure of strongly interacting matter at large densities \cite{Brandes2024}.
Our approach is conducive to the use of a more complex EoS parameterization.

\section{Conclusion}
\label{sec:conclusion}

Neural likelihood estimation (NLE) and Hamiltonian Monte Carlo (HMC) are used in conjunction to allow the first inference of the full posterior for neutron star equation of state and nuisance parameters directly from telescope spectra. In realistic scenarios for nuisance parameter priors, this method outperforms all state-of-the-art methods in terms of prediction accuracy. 

Extracting these parameters requires simulation-based inference techniques, as the likelihood is analytically unavailable. Previous methods either relied on inference of the mass and radius as an intermediate step or only provided a point estimate of the EoS parameters. NLE+HMC allows for the full posterior, avoiding simplifying assumptions and information collapse at an intermediate step, and is trivially extendable to multiple stars and other data.

Our study considered three different scenarios for the available prior information on the nuisance parameters coming from additional observations beyond the telescope spectra. With tighter prior constraints, the marginal posterior distributions of the EoS parameters, and accordingly the posterior $P(\varepsilon)$ and $R(M)$ credible bands, become much tighter. Both the hydrogen column $N_H$ and effective surface temperature $\log(T_\mathrm{eff})$ can be tightly constrained from the spectra. For the distance $d$, the posterior constraints are driven by the prior information.
    
These techniques can be extended to NICER or gravitational wave data \cite{Goncalves2023}\footnote{For the analysis of NICER and gravitational waves, the telescope data is much larger compared to the quiescent LMXBs analyzed here. In that case, it might become necessary to use an additional embedding net to learn summary statistics \cite{TejeroCantero2020}.}. Many more gravitational wave events from binary neutron star mergers will be detected in future runs of LIGO, Virgo, and KAGRA. In addition, the gravitational wave database will increase dramatically with the next generation of gravitational wave detectors. For such a large number of expected observations, it is not numerically feasible to compute the likelihood by integrating KDEs \cite{Lange2018,Vivanco2019,Krastev2023}. Neural simulation-based inference techniques promise to provide a cost-efficient alternative~\cite{Dax2021,McGinn2024}. 
    
This approach can help guide decisions about future observations. Sec.~\ref{sec:observationnumber} demonstrated our method can easily be used to study the impact of additional measurements on the final EoS constraints. It can therefore be used to estimate the relative value of future measurements of one star compared to another, using simulations. Through the assessment of the constraining power of multiple simulated test spectra, we can anticipate which stars provide the most useful additional information required to further constrain the equation of state, thereby guiding decisions for future experimental endeavors.

Based on the results derived in this study on simulated test data, we can conclude that neural likelihood estimation techniques provide a promising avenue for the inference of the neutron star equation of state, not only for X-ray measurements of neutron stars but possibly also for gravitational waves. In future studies, this method would be applied to real telescope measurements to draw conclusions about the internal structure of neutron stars. 

\section*{Acknowledgements}

LB thanks Max Dax, Kenji Fukushima, and Koichi Murase for stimulating discussions. LB is supported in part by Deutsche Forschungsgemeinschaft (DFG) and the National Natural Science Foundation of China (NSFC) through funds provided by the Sino-German CRC110 (DFG Grant No. TRR110 and NSFC Grant No. 11621131001), and by the DFG Excellence Cluster ORIGINS. LB thanks the Japan Society for the Promotion of Science (JSPS) and The University of Tokyo for their support and hospitality during his International Fellowship for Research in Japan in the winter and spring of 2024. AG and DW are supported by The Department of Energy Office of Science. FW and DF are supported by the National Science Foundation (USA) under Grant No. PHY-2012152. AWS is supported by NSF AST NSF grant, AST 22-06322 and PHY 21-16686. LL is supported by NSF grant 2012857. LH is supported by the Excellence Cluster ORIGINS, which is funded by the Deutsche Forschungsgemeinschaft (DFG, German Research Foundation) under Germany’s Excellence Strategy - EXC-2094-390783311.

AG used resources from the National Energy Research Scientific Computing Center (NERSC), a U.S. Department of Energy Office of Science User Facility located at Lawrence Berkeley National Laboratory, for this research. We specifically thank Wahid Bhimji and Benjamin Nachman for their support with these computing resources. This collaboration was forged in discussions at the Differentiable and Probabilistic Programming for Fundamental Physics Workshop at Munich and the research was therefore supported by the Munich Institute for Astro-, Particle and BioPhysics (MIAPbP) which is funded by the DFG under Germany's Excellence Strategy -- EXC-2094 -- 390783311.

\appendix
\section{Relativistic mean-field model}
\label{sec:RMF}

In this study, the EoS is based on a relativistic mean-field (RMF) model, which expresses the baryon-baryon interactions through effective meson exchanges. These mesons encompass a scalar meson ($\sigma$) signifying attraction between baryons, a vector meson ($\omega$) representing repulsion, and an isovector meson ($\rho$) essential for elucidating nucleon-nucleon interactions in isospin asymmetric systems. The pion, playing a key role in the long-range description of baryon-baryon interaction, assumes odd parity and consequently disappears in the mean-field approximation.
The Lagrangian governing interacting nucleons $\psi_N$ is expressed as \cite{glendenning2012compact}:
\begin{equation} \label{eq:lagrangian-baryons}
  \begin{aligned}
    \mathcal{L}_{\mathrm{N}} = \sum\limits_N
{\bar\psi}_N\bigl[\gamma_{\mu}\bigl(i\partial^{\mu}
      -g_{\omega N} \omega^{\mu}-\tfrac{1}{2}g_{\rho N} \, \boldsymbol{\tau}\cdot
    \boldsymbol{\rho}^{\mu}\bigr)
    -\bigl(m_N-g_{\sigma N} \, \sigma\bigr)\bigr]\psi_N \, ,
  \end{aligned}
\end{equation}
where $g_{\sigma N}$, $g_{\omega N}$, and $g_{\rho N}$ represent the meson-nucleon coupling constants.

In the standard RMF approach, meson-nucleon coupling constants are set to reproduce properties of isospin symmetric nuclear matter at nuclear saturation density, $n_0$. These properties include the energy per nucleon $E_0$, the nuclear incompressibility $K_0$, the effective nucleon mass $m^*/m_N$, the asymmetry energy $J$, the asymmetry energy slope $L_0$, and the nucleon potential $U_N$. For the GM1L parameter set employed in this study, the respective values are $n_0 = 0.153\,$fm$^{-3}$, $E_0 =-16.3\,$MeV, $K_0 = 300 \,$MeV, $m^*/m_N = 0.70$, $J = 32.5\,$MeV, $L_0 = 55.0\,$MeV, and $U_N = -65.5\,$MeV.
Electrons and muons in neutron star matter are described by the lepton Lagrangian ($\lambda \in \{e^-,\mu^-\}$):
\begin{equation} \label{eq:lagrangian-leptons}
  \mathcal{L}_{\mathrm{L}} = \sum_{\lambda}
  \overline\psi_{\lambda}\bigl(i\gamma_{\mu}
    \partial^{\mu}-m_\lambda\bigr)\psi_{\lambda} \, ,
\end{equation}
while the meson Lagrangian is given by \cite{Gambhir1990, glendenning2012compact}:
\begin{equation} \label{eq:lagrangian-mesons}
    \mathcal{L}_{\mathrm{M}} =  
    \, \tfrac{1}{2}\bigl(\partial_{\mu}\sigma\partial^{\mu}\sigma-
    m^2_{\sigma}\sigma^2\bigr)-\tfrac{1}{4}\omega_{\mu\nu}\omega^{\mu\nu}
    +\tfrac{1}{2}m^2_{\omega}\omega_{\mu}\omega^{\mu}\\
    +\tfrac{1}{2}m^2_{\rho}\boldsymbol{\rho}_{\mu}\cdot
    \boldsymbol{\rho}^{\mu}-\tfrac{1}{4}\boldsymbol{\rho}_{\mu\nu}\cdot
    \boldsymbol{\rho}^{\mu\nu}~,
\end{equation}
where $\omega_{\mu\nu} = \partial_{\mu}\omega_{\nu}-\partial_{\nu}\omega_{\mu}$ and $\boldsymbol\rho_{\mu\nu} = \partial_{\mu}\boldsymbol\rho_{\nu}-
\partial_{\nu}\boldsymbol\rho_{\mu}$. To ensure that the RMF model reproduces the empirical values for the nuclear incompressibility and the effective nucleon mass at saturation, additional nonlinear scalar self-interactions need to be included in the Lagrangian \cite{Boguta1977}:
\begin{equation} \label{eq:lagrangian-scalar-selfinteractions}
  \begin{aligned}
    \mathcal{L}_{\mathrm{NL\sigma}} = -\tfrac{1}{3}b_{\sigma}m_N
    \bigl[g_{\sigma N}(n)\sigma\bigr]^3 -\tfrac{1}{4}c_{\sigma}
    \bigl[g_{\sigma N}(n)\sigma\bigr]^4~,
  \end{aligned}
\end{equation}
where $b_{\sigma}$ and $c_{\sigma}$ are constants determined by the properties of symmetric nuclear matter.

The field equations resulting from the aforementioned Lagrangians must be solved while adhering to the conditions of electrical charge neutrality in neutron star matter:
\begin{equation} \label{eq:charge-neutrality}
  n \, q_p
  +\sum\limits_{\lambda} n_{\lambda} q_{\lambda}  = 0 ~,
\end{equation}
where $q_p$ is the electric charge of a proton, 
and conservation of baryon number:
\begin{equation} \label{eq:baryon-number-conservation}
  n - \sum\limits_{B=n, p}  n_B =0 ~.
\end{equation}
The meson-field equations combined with the 
equations for electric charge neutrality and 
baryon number conservation constitutes a set of 
five coupled nonlinear equations, to be simultaneously solved to determine the meson mean-fields ($\bar\sigma$, $\bar\omega$, $\bar\rho$) and the neutron and electron Fermi momenta ($k_n$, $k_e$). The proton's Fermi momentum is determined by the condition that neutron star matter is in chemical equilibrium:
\begin{equation} \label{eq:chemical-equilibrium}
  \mu_p = \mu_n - q_p \mu_e ~,
\end{equation}
where $\mu_n$ and $\mu_p$ are the chemical potentials of neutrons and protons respectively, and $\mu_e$ is the electron chemical potential. The presence of muons in neutron star matter occurs when $\mu_e \geq \mu_\mu$. For the neutron star matter EoS, we match the GM1L parametrization with the Baym-Pethick-Sutherland (BPS) model \cite{Baym1971} for the outer crust and the Baym-Bethe-Pethick (BBP) model \cite{Baym1971b} for the inner crust. 

To limit the number of parameters in the inference procedure, it is advantageous to represent the EoS with only a few parameters. As described in Ref.~\cite{Farrell2023a}, we accomplish this by constructing a parametric representation based on a spectral fit, where the EoS is represented as a linear combination of basis functions and can therefore be reconstructed using the coefficients of the basis functions. Using a second-order expansion, we construct a spectral fit of the GM1L EoS using the process outlined in Refs. \cite{Lindblom2010, Lindblom2018}. The two coefficients from the spectral fit are hereafter referred to as $\lambda_1$ and $\lambda_2$. The original coefficients describing the GM1L EoS are then used to generate many EoS scenarios using the expression:
\begin{align}
    \lambda_{\text{generated}} = \lambda_{\mathrm{fit}} \cdot(1+2 \cdot c \, (\mathit{ran2}-0.5))
\end{align}
where $\lambda_{\text{generated}}$ represents the newly constructed spectral parameter, $\lambda_{\text{fit}}$ is the best fit spectral parameter of GM1L, and $c$ is a scaling parameter set to 0.05.  $\mathit{ran2}$ are uniformly distributed random numbers in the range 0 to 1 generated by the $\mathit{ran2}$ function given in \cite{Press2007}. This process was repeated to create around $10^3$ different EoS models, each represented by a unique set of spectral coefficients $\lambda_1$ and $\lambda_2$. Following this process, the two parameters are uniformly distributed in the intervals $\lambda_1 \in [4.75, 5.25]$ and $\lambda_2 \in [-2.05, -1.85]$.

\section{Normalizing flows and simulation-based inference}
\label{sec:NormalizingFLow}
Normalizing Flows are a class of generative models in machine learning that focus on learning a bijective mapping between a simple base distribution $\pi(u)$ (usually chosen to be a Gaussian distribution) and a more complex, target distribution $p(x)$ \cite{Papamakarios2017a}. The main idea is to model $p(x)$ by transforming the base distribution $\pi(u)$ through a series of invertible and differentiable transformations $f_{\Phi}$ with trainable parameters $\Phi$. %\textcolor{highlight}{These transformations are volume preserving by design~\footnote{\textcolor{highlight}{Although there exist certain non-volume preserving variants.}}.}

Given $N$ transformations, $f_{\Phi} = f_{\Phi_1} \circ \dots \circ f_{\Phi_N}$, we can easily generate samples $x$ from $p(x)$ by transforming the samples $u$ of $\pi(u)$
\begin{align}
    &u \sim \pi(u) ~, \\
    &x = f_{\Phi}(u) ~.
\end{align}
From the change of variables law for probability distributions, we can compute the probability density of the target distribution 
\begin{align}
    p(x) = \pi(f^{-1}_{\Phi}(x)) \bigg|\det \left(\frac{\partial f^{-1}_{\Phi}}{\partial x} \right)\bigg| ~.
\end{align}
Consequently, for a fast numerical evaluation of the probability density, the transformations $f_{\Phi}$ should be chosen such that they are easy to invert and have a Jacobian whose determinant should be fast to compute. One popular choice is Masked Autoregressive Flows (MAF) \cite{Papamakarios2017b}, where each dimension of the data is sequentially transformed conditioned on the previous dimensions, making the Jacobian of $f_\Phi^{-1}$ triangular by design. MAFs are well suited for our purposes because they are very efficient in computing the probability density $p(x)$, but less efficient in generating samples, $x \sim p(x)$.

%Normalizing flows have found applications in various domains, including generative modeling and density estimation, due to their ability to model complex distributions and generate new samples from them.

In simulation-based inference (SBI), our goal is to train a conditional normalizing flow to learn the likelihood distribution, $p(s|\lambda,\nu)$.
This is achieved my minimizing the  Kullback-Leibler divergence ($D_\text{KL}$), which is a measure of the statistical distance between two probability distributions, between the likelihood $p(s|\lambda,\nu)$ and the distribution parameterized by the normalizing flow, $q_\Phi(s|\lambda,\nu)$.  $D_\text{KL}$ becomes zero if the distributions are identical. Hence we fit the  trainable parameters $\Phi$ with the following optimization procedure:
% The Kullback-Leibler divergence, $D_\text{KL}$, is a measure of the statistical distance between two probability distributions, i.e., it becomes zero if the distributions are identical. 
% Consequently, to approximate a likelihood distribution, $p(s|\lambda,\nu)$, we can train a normalizing flow, $q_\Phi(s|\lambda,\nu)$ (with trainable parameters $\Phi$), to minimize the Kullback-Leibler divergence between the likelihood and the approximate distribution
\begin{align}
     \arg\min_{\Phi}  D_\text{KL} \left(p(s|\lambda,\nu) \big|\big|q_\Phi(s|\lambda,\nu) \right) &=  \arg\min_{\Phi}  \int \mathrm{d}s\,  p(s|\lambda, \nu) \, [\log p(s|\lambda,\nu)  - \log q_\Phi(s|\lambda,\nu)] \nonumber \\
     &\approx \arg \min_{\Phi}  \sum_{s_i\sim p(s|\lambda_i, \nu_i)} \log p(s_i|\lambda_i,\nu_i)  - \log q_\Phi(s_i|\lambda_i,\nu_i) \nonumber \\
     &= \arg \min_{\Phi}  \sum_{s_i\sim p(s|\lambda_i, \nu_i)}  - \log q_\Phi(s_i|\lambda_i,\nu_i) \nonumber \\
     &= \arg \max_{\Phi}  \sum_{s_i\sim p(s|\lambda_i, \nu_i)}  \log q_\Phi(s_i|\lambda_i,\nu_i) ~,
\end{align}
where the second line is the Monte-Carlo estimator of the integral in the definition of the KL divergence in the first line. The first term in the second line can be dropped as it is constant with respect to the parameters $\Phi$ of the density estimator. Thus to learn the likelihood distribution, training the density estimator to minimize the KL divergence is equivalent to maximizing the log-probability of the sampled spectra, $s_i$, generated from the likelihood distribution via the simulator for given EoS and nuisance parameters, $(\lambda_i, \nu_i)$.

\section{Hamiltonian Monte Carlo}
\label{sec:HMC}

Let $x\in \mathcal{R}^n$ denote the state (random variable) in a continuous state space. 
% For HMC sampling of a target density over $\R^d$, we will have $n=2d$ after coupling momentum with position.
Our goal is to generate samples from a target probability distribution function $\pi(x)$.
We assume the normalization $\int dx\, \pi(x) = 1$, 
although all MCMC methods only require un-normalized densities.

\subsection{Metropolis-Hastings}
\label{s:mh}

Given the target distribution $\pi$ and the current state $x$,
the random-walk Metropolis-Hastings algorithm constructs a Markov chain by sampling a proposal $y$ from a transition function (proposal distribution) $t(x, y)$.
The simplest example of such a proposal distribution is a Gaussian distribution centered on $x$ with some width $\sigma$, i.e., $y~\sim \mathcal{N}(x, \sigma^2)$.
The proposal is then accepted with some probability $\alpha(x,y)$, in which case the next state is $y$, otherwise it remains $x$.
We need to identify this acceptance probability to ensure that asymptotically this Markov chain generates samples from the target distribution, $x,y\sim \pi$. This is equivalent to ensuring that the target distribution $\pi$ is invariant under this Markov chain, for example by maintaining detailed balance conditions when accepting the proposed stage 
\begin{equation}
\pi(x) t(x,y) \alpha(x,y) = \pi(y) t(y,x) \alpha(y,x) ~.
\label{eq:alrat}
\end{equation}
Detailed balance means ensuring that the function $\pi(x)t(x,y)\alpha(x,y)$
is symmetric in exchanging $x$ and $y$ ($x\leftrightarrow y$).  
This equation enforces that the probability of being in the state $x$, proposing a transition to state $y$ and accepting this transition ($x\rightarrow y$) is the same as making the reverse transition ($y \rightarrow x$).
% going forward $x\rightarrow y$ and backward $y \rightarrow x$.

If $t(x,y)$ is positive everywhere, the above condition is satisfied by the standard Metropolis-Hastings acceptance formula for $x, y \in S$,
\begin{equation}
\alpha(x,y) = \min\!\left(\frac{\pi(y) \, t(y,x)}{\pi(x) \, t(x,y)}, 1 \right),
\label{eq:al}
\end{equation}
where the denominator is never zero given the above assumptions on $\pi$ and $t$.
For each $x,y\in \mathcal{R}^n$,
either $\alpha(x,y)$ or $\alpha(y,x)$ is 1. There are other formulae for $\alpha$ obeying Eq.\,\,\eqref{eq:alrat},
but with lower acceptance rates, meaning they lead to an undesirable higher asymptotic variance for estimated expectations.

\subsection{Classical Hamiltonian Monte Carlo}
\label{s:hmc}

In high dimensions, random walk Metropolis-Hastings as described in the previous section often leads to diffusive behavior and can be extremely inefficient. In this section, we outline the Hamiltonian Monte Carlo (HMC) sampling algorithm which overcomes this by designing more efficient proposal kernels, $t(x,y)$, that utilize gradient information~\cite{Neal2011,Betancourt2017}.
Following standard notation for Hamiltonian dynamics, $q\in \mathcal{R}^n$ denotes the parameter of interest that is to be sampled.
The target density, $\pi$, is assumed to be continuous, differentiable, and positive everywhere.
To draw samples $q$ from $\pi(q)$, HMC reinterprets the parameters of interest as
a position vector with associated potential energy function $U(q) = -\log \pi(q)$.
%and simulates a Markov chain by approximating the following Hamiltonian dynamics.
We introduce an auxiliary momentum vector $p\in \mathcal{R}^n$, which contributes a kinetic energy term $K(p) = \frac{1}{2}p^T M^{-1}p$, where $M$ is some symmetric positive definite mass matrix
that we take as fixed.
Then the Hamiltonian $H:\mathcal{R}^{2n} \to \mathcal{R}$ gives the total energy for the state $x := (q,p)$,
\begin{equation}
    H(x) = H(q,p) = U(q) + \frac{1}{2}p^T M^{-1} p ~.
\label{H}
\end{equation}
The state space $S = \mathcal{R}^{2n}$ is called \textit{phase space}. The dynamical evolution of particles under this Hamiltonian is called Hamiltonian flow and can be simulated by solving Hamiltonian equations.

HMC uses a Markov chain to generate samples $x$ from the canonical distribution $\tilde{\pi}$ defined by $H$, namely
\begin{equation}
\tilde \pi(x) := Z^{-1} \, e^{-H(x)} = Z^{-1} \, e^{-U(q)} \, e^{-\frac{1}{2} p^T M^{-1} p} = Z^{-1} \, \pi(q) \, e^{-\frac{1}{2} p^T M^{-1} p} ~,
\label{eq:Gibbs}
\end{equation}
where $Z=\int_{\mathcal{R}^{2n}} dx\, e^{-H(x)} = (2\pi)^{n/2}\sqrt{\det M}$ is a normalizing constant.
Since $H$ is the sum of potential and kinetic terms, in the Gibbs density
$q$ and $p$ are independent, with
the $q$-marginal of $\tilde \pi(x)$ being the target density $\pi(q)$.
Thus, extracting the first $n$ coordinates of samples $x^{(i)}$ from $\tilde 
 \pi$, one obtains samples from $\pi$.

HMC constructs a Markov chain to generate samples from this distribution.
Given a current state $x^{(i)}:=(q^{(i)},p^{(i)})$, the Markov update comprises two steps:
\begin{description}
    \item[Step 1. Gibbs sampling:] Resample the momentum $p^{(i)}$ from its Gaussian marginal distribution $p \sim \mathcal{N}(0, M)$, without changing $q^{(i)}$. This randomization step is needed to mix efficiently between different $H$ values (corresponding to energy level-sets). 
    
    \item[Step 2. Metropolis update:] Given the momentum $p^{(i)}$,
    generate a new Metropolis-Hastings proposal via a deterministic map $y=F(x)$ which approximates the Hamiltonian flow in Eq.\,\,\eqref{H} over a certain time $T$, starting from the initial point $x=(q^{(i)},p^{(i)})$ and is followed by negation of the final momentum\footnote{This negation of momentum maintains the invertibility of every step as is necessary for detailed balance, but is not necessary in practice for HMC as it is followed with a Gibbs momentum refresh step.}. This map defines the transition kernel $t$.
    This proposal is then accepted with the probability $\alpha(x,y)$. 
 \end{description}

The most commonly used dynamics for the map $F$ approximating the Hamiltonian flow, i.e., solving the Hamiltonian equations, is the leapfrog (Verlet) integrator.
Each leapfrog step, written $(q',p') = L_\epsilon(q,p)$, comprises three substeps:
 \begin{align}
     \bar{p}     &\leftarrow\; p - \frac{\epsilon}{2} \nabla U(q)~,\nonumber \\
     q' &\leftarrow\; q + \epsilon M^{-1} \bar{p}~, \nonumber \\
     p' &\leftarrow\; \bar{p} - \frac{\epsilon}{2} \nabla U(q') ~,
     \label{Leps}
 \end{align}
where $\epsilon$ is the step-size. 
We compose $n = T/\epsilon$ such steps, $(L_\epsilon)^n$ to integrate the Hamiltonian flow for time $T$.
% , is a $\bigO(\epsilon^2)$-accurate approximation to Hamiltonian dynamics for time $T$ \cite{leimkuhler2004simulating}.
With the momentum-flip operator $P(q,p) := (q,-p)$, $F=(L_\epsilon)^n P$
(operators act left to right)
is volume-preserving because each of the three substeps is a shear transformation\footnote{A \textit{shear} is a map of the form $(q,p)\mapsto(q+G(p),p)$ or $(q,p+G(q))$, and
it is easy to check that the determinant of the $2n\times2n$ Jacobian derivative matrix is 1.},
and $F$ is an involution\footnote{Involution means $F^{-1}=F$, i.e., it is \textit{time reversible}} because $L_\epsilon(q',-p')=(q,-p)$, which can be verified by reversing the order of the substeps, so $L_\epsilon P L_\epsilon = P$ and $((L_\epsilon)^n P)^2 = I$.
In this case, the transition described in Step~2 preserves detailed balance when
\begin{equation}
\alpha(x,y) = \min\left(\frac{\tilde\pi(y)}{\tilde\pi(x)}, \, 1\right),
\label{eq:alhmc}
\end{equation}
so that the distribution $\tilde \pi$ is invariant under Step 2.
Since $\tilde \pi$ is also invariant under the Gibbs sampling in  Step 1,
$\tilde \pi$ is invariant under their composition of both steps, i.e., under each
HMC update.
Note that failure to approximate well the Hamiltonian flow by the leapfrog integrator
\textit{does not} impact detailed balance,
although it can drastically reduce the mixing of the Markov chain, and hence the efficiency of the algorithm.

\subsection{Implementation}

To draw posterior samples with HMC, we run 16 chains of 2000 samples each. The leapfrog integration in each chain is performed for 40 steps to generate new proposals, with a step-size that is dynamically determined using dual averaging for 300 burn-in steps to achieve an average acceptance probability of 0.65.  The mass matrix used in this work is diagonal, except for negative off-diagonal elements between the EoS parameters. In this work, we tuned these values manually based on few initial runs, but in the future this can be automated using variational approximations to the target distribution~\cite{Modi2023b}. We use an importance sampling step to determine initial values for each chain \cite{Lueckmann2021}. To facilitate an easy comparison of our
NLE + HMC method with future analyses, we provide working examples of the algorithms implemented for this work in a public GitHub repository\footnote{\url{https://github.com/lenjonah/neutron_star_inference}}.  

%\clearpage

%\iffalse

\bibliography{ns,baldi,nle_library}

\end{document}